\documentclass[11pt, a4paper]{article}

\usepackage[utf8]{inputenc} % For handling UTF-8 text
\usepackage[T1]{fontenc}    % For improved font encoding
\usepackage{geometry}       % For setting page margins
\usepackage{graphicx}       % For including images
\usepackage{amsmath}        % For advanced math environments
\usepackage{amssymb}        % For math symbols
\usepackage{booktabs}       % For professional-looking tables
\usepackage{url}            % For formatting URLs
\usepackage{listings}       % For code snippets
\usepackage{xcolor}         % For defining colors
\usepackage[font=footnotesize]{caption}        % For customizing captions
\usepackage{subcaption}     % For subfigures
\usepackage{hyperref}       % For clickable links and references

\newcommand{\datasetname}{LOOPerSet}
\hypersetup{
    colorlinks=true,
    linkcolor=blue!70!black,  % Color for internal links
    citecolor=green!60!black, % Color for citations
    urlcolor=purple!80!black, % Color for URLs
    pdftitle={\datasetname{}: A Large-Scale Dataset for Data-Driven Polyhedral Compiler Optimization},
    pdfauthor={Your Name(s) Here},
}
\hypersetup{pdfcreator={MyName},pdfproducer={MyTool}}

\definecolor{codegreen}{rgb}{0,0.6,0}
\definecolor{codegray}{rgb}{0.5,0.5,0.5}
\definecolor{codepurple}{rgb}{0.58,0,0.82}
\definecolor{backcolour}{rgb}{0.95,0.95,0.95}

\lstdefinestyle{jsonstyle}{
    backgroundcolor=\color{backcolour},   
    commentstyle=\color{codegray},
    keywordstyle=\color{magenta},
    numberstyle=\tiny\color{codegray},
    stringstyle=\color{codepurple},
    basicstyle=\ttfamily\scriptsize,
    breakatwhitespace=false,         
    breaklines=true,                 
    captionpos=b,                    
    keepspaces=true,                 
    numbers=left,                    
    numbersep=5pt,                  
    showspaces=false,                
    showstringspaces=false,
    showtabs=false,                  
    tabsize=2
}
\definecolor{eclipseStrings}{RGB}{42,0.0,255}
\definecolor{eclipseKeywords}{RGB}{127,0,85}
\colorlet{numb}{magenta!60!black}

\lstdefinelanguage{json}{
    basicstyle=\scriptsize\ttfamily,
    commentstyle=\color{eclipseStrings}, % style of comment
    stringstyle=\color{eclipseKeywords}, % style of strings
    numbers=left,
    numberstyle=\scriptsize,
    stepnumber=1,
    numbersep=8pt,
    showstringspaces=false,
    breaklines=true,
    frame=lines,
    backgroundcolor=\color{gray}, %only if you like
    string=[s]{"}{"},
    comment=[l]{//},
    morecomment=[s]{/*}{*/},
    literate=
        *{0}{{{\color{numb}0}}}{1}
         {1}{{{\color{numb}1}}}{1}
         {2}{{{\color{numb}2}}}{1}
         {3}{{{\color{numb}3}}}{1}
         {4}{{{\color{numb}4}}}{1}
         {5}{{{\color{numb}5}}}{1}
         {6}{{{\color{numb}6}}}{1}
         {7}{{{\color{numb}7}}}{1}
         {8}{{{\color{numb}8}}}{1}
         {9}{{{\color{numb}9}}}{1}
}

\title{\textbf{\datasetname{}: A Large-Scale Dataset for Data-Driven Polyhedral Compiler Optimization}}

\author{
    Massinissa Merouani\textsuperscript{1} \and
    Afif Boudaoud\textsuperscript{1}\thanks{Currently at ETH Zurich.} \and
    Riyadh Baghdadi\textsuperscript{1}
}
\date{%
    \small
    \textsuperscript{1}New York University Abu Dhabi
}

\begin{document}

\maketitle

\begin{abstract}
\noindent % No indentation for the abstract
The advancement of machine learning for compiler optimization, particularly within the polyhedral model, is constrained by the scarcity of large-scale, public performance datasets. This data bottleneck forces researchers to undertake costly data generation campaigns, slowing down innovation and hindering reproducible research learned code optimization. To address this gap, we introduce \datasetname{}, a new public dataset containing 28 million labeled data points derived from 220,000 unique, synthetically generated polyhedral programs. Each data point maps a program and a complex sequence of semantics-preserving transformations—such as fusion, skewing, tiling, and parallelism—to a ground-truth performance measurement (execution time). The scale and diversity of \datasetname{} make it a valuable resource for training and evaluating learned cost models, benchmarking new model architectures, and exploring the frontiers of automated polyhedral scheduling. The dataset is released under a permissive license to foster reproducible research and lower the barrier to entry for data-driven compiler optimization.
\end{abstract}

%==============================================================================
\section{Introduction}

The polyhedral model~\cite{Feautrier2011,thangamani_survey_2024} offers a powerful framework for expressing and applying complex loop transformations, which are essential for optimizing scientific and high-performance applications. The central challenge, however, lies in navigating the vast search space of legal transformation sequences to find one that delivers optimal performance on a given hardware target. For years, this complex search has been effectively guided by established analytical cost models and sophisticated heuristics~\cite{10.1145/2896389,bondhugula_practical_2008,grosser_polly_2012,pop_graphite_2006,pouchet_polyopt_2012}. These methods are designed to be general and tractable, typically optimizing for key performance proxies like improving data locality and enabling parallelism.

A different and complementary approach is to replace these analytical models with empirical ones built using machine learning. By training on large amounts of performance data, a learned model can develop a more nuanced understanding of the true cost and benefit of transformations on real hardware. It can learn to capture the subtle, non-linear interactions between optimizations and the underlying system that are difficult to model analytically. This data-driven approach, validated by prior work like the Tiramisu autoscheduler~\cite{baghdadi_deep_2021} and others~\cite{chen_learning_2018,zheng_ansor_2020,adams_learning_2019,phothilimthana_tpugraphs_2023}, has shown great potential for discovering highly effective optimization strategies.

However, the efficacy of this data-driven approach is constrained by the limited availability of large-scale, public performance datasets for polyhedral optimization. This data scarcity presents a significant obstacle to progress, as it necessitates that individual research groups undertake their own expensive and time-consuming data generation campaigns. This practice not only slows down the pace of innovation but also hinders reproducibility and makes it difficult to perform rigorous comparisons between new data-driven approaches. It raises the barrier to entry, discouraging researchers who could otherwise contribute novel machine learning techniques to the field.

To address this limitation, we introduce  \datasetname{}, a large-scale public dataset created to facilitate research in data-driven polyhedral compilation.  \datasetname{}  consists of over 28 million labeled data points derived from a diverse corpus of approximately 220,000 unique, synthetically generated polyhedral programs. Each data point maps a rich program representation and a legal transformation schedule to a ground-truth performance label measured on physical hardware. The dataset was built with two key principles: maximizing the structural diversity of its programs and ensuring the relevance of its sampled transformations. Our goal is to provide a foundational artifact that allows the community to focus on what matters most: designing, training, and evaluating the next generation of learned optimizers.

The dataset was originally created to develop and train the LOOPer autoscheduler~\cite{looper_pact25}. While that paper details the architecture and evaluation of the LOOPer system, this paper serves as the reference for the dataset itself. We begin by providing a high-level overview of the dataset in Section 2. We then detail our relevance-guided data generation process in Section 3 and present a rigorous statistical characterization and quantitative diversity analysis in Section 4 to validate the dataset's quality. Finally, Section 6 provides instructions on the dataset's format, licensing, and access to enable its immediate use by the research community.

%==============================================================================
\section{Dataset Overview}

\datasetname{} is a large-scale public dataset intended to support research and development in data-driven code optimization. At its core, the dataset is a collection of examples where each data point maps a program and a specific sequence of optimizations to a measured, ground-truth performance speedup. In total, it comprises over 28 million labeled examples derived from a corpus of approximately 220,000 unique programs.

The base programs used in the \datasetname{} were generated synthetically while ensuring the coverage of a wide and diverse range of computational patterns and structures, extending beyond what is typically found in standard benchmark suites. Every program uses static affine control, making it fully analyzable within the polyhedral model.

For each program, the dataset contains performance data for numerous optimization sequences. These sequences are composed of key polyhedral and classical transformations, such as loop fusion, skewing, interchange, reversal, tiling, parallelization, and unrolling. Every transformation sequence applied was first checked for correctness using polyhedral dependency analysis, guaranteeing that it preserves the original program's semantics.

The performance of each optimized program variant was measured directly on physical hardware to provide a ground-truth label. This label is obtained by measuring execution time from many separate runs to ensure our measurements are stable and reliable. %All measurements were collected on a dual-socket server equipped with Intel Xeon E5-2695 v2 processors. 

The dataset is designed to be a versatile tool for the compiler community. Its most direct use is as a large-scale benchmark for training and evaluating machine learning models on code performance related tasks. Researchers can also analyze the data directly to discover new, data-driven compiler heuristics. Finally, it provides a foundation for tackling the hardware portability problem, enabling models to be pre-trained and subsequently fine-tuned for a new architecture with substantially reduced data collection requirements.

The full dataset is publicly available and free to use for academic and commercial research. It is accessible on Hugging Face\footnote{\url{https://huggingface.co/datasets/Mascinissa/LOOPerSet}} along with tools and examples for parsing the data. Researchers can use either the full 28-million-point dataset or a smaller 10-million-point version that was used for the experiments in the LOOPer paper~\cite{looper_pact25}.

%==============================================================================
\begin{figure}[h!]
    \centering
    \includegraphics[width=\textwidth]{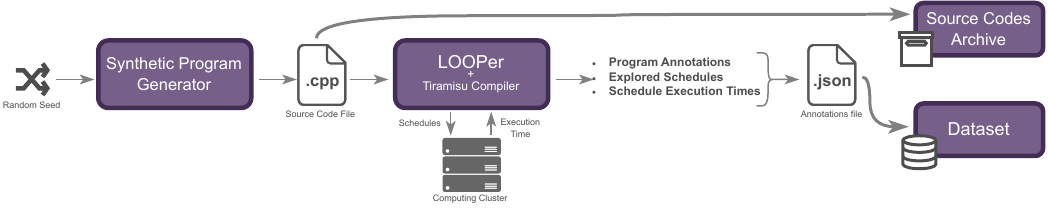} 
    \caption{The end-to-end data generation pipeline.}
    \label{fig:pipeline}
\end{figure}
\section{Data Generation Methodology}

Our dataset was built using a unified pipeline composed of the three main stages described below: program generation, transformation sampling, and performance labeling. As illustrated in Figure~\ref{fig:pipeline}, this process was designed to produce a collection of structured JSON files which form the final dataset.

For each synthetic program, the pipeline first extracts a comprehensive set of features, everything needed to fully describe the program, including its loop structure, iteration bounds, memory access patterns, computational expressions, etc. These features are recorded in a dedicated JSON object. Then, for every transformation schedule we explored for that program, the schedule features and its measured execution time are appended to that same JSON object. The result is a set of structured machine-readable files where each file contains one program and a list of all evaluated optimization schedules, along with their corresponding performance. The individual JSON files are then merged into a single dataset file. The detailed format of these JSON files is described later in Section~\ref{sec:format}.

The following subsections describe each stage of this generation process in detail.

\subsection{Program Space Sampling: The Synthetic Program Generator}

To build a dataset that is both large and diverse, we developed a synthetic program generator. Our goal was to create a wide variety of program structures and computational patterns to better reflect the complexity of real-world scientific code. To achieve this, our generator follows a multi-stage randomized process designed to maximize structural diversity and avoid biasing towards any specific benchmark.

At its core, the generation process builds a program recursively, starting with its overall structure and progressively adding detail. The key stages are:

\paragraph{Loop Structure Generation:} The process begins by creating the program's loop skeleton. The generator probabilistically decides on the number of top-level loop nests and then, for each nest, recursively builds out its structure. At each level, it probabilistically decides whether to create further nested loops or to terminate the branch. This method produces a wide variety of program shapes, from simple, deep nests to complex programs with multiple, shallow loop sequences. To enhance structural diversity, the generator creates both rectangular and non-rectangular (e.g., triangular, trapezoidal) iteration domains, with loop bounds that can be either constants or functions of outer loop iterators.
  
\paragraph{Computation Placement and Ordering:} Once a loop structure is formed, the generator populates it by randomly placing computations within the nests. It can intersperse computations with sub-nests at the same level, creating interleaved program structures. Each computation is also assigned a data type (e.g., 32/64-bit floats or integers) from a predefined distribution.
    
\paragraph{Memory Access and Expression Generation:} This stage defines computational workload of each computation. At this stage, the generator shapes the overall arithmetic expression performed by the computation and decides what data it accesses. 
\begin{itemize}
    \item \textbf{Memory Patterns:}  For each computation, the generator decides which data arrays (buffers) it reads from and writes to. This process can reuse existing buffers to create data dependencies or introduce new buffers that act as program inputs. Importantly, the generator creates a variety of memory access patterns, from simple identity mappings to more complex idioms like multi-dimensional stencils (e.g., star-shaped or cross-shaped) and accesses with  constant offsets.
     \item \textbf{Arithmetic Expressions:} The computational logic itself is built by randomly composing an expression tree. This tree combines the previously defined memory accesses with scalar values using a mix of common arithmetic operators (\texttt{+},  \texttt{-},  \texttt{*},  \texttt{/}) and mathematical functions (\texttt{sqrt},  \texttt{max},  \texttt{min}). This results in programs with a wide range of operational intensities and complexities.
\end{itemize}

\paragraph{Coherence and Validation Pass:} A purely random process can generate nonsensical programs. To ensure that our synthetic programs are not just syntactically valid but also computationally meaningful, the generator performs a final coherence check. After a computation is formed, it is analyzed to detect and prevent trivial work. For instance, it checks if a loop's iterations are independent of the final result (computationally redundant loops), if a memory location is overwritten without an intermediate read (dead writes), etc. If such an issue is found, the computation is discarded and regenerated until it represents meaningful work.

This multi-stage process, combining randomized structural generation with a final coherence pass, allows us to effectively and efficiently sample a vast and diverse region of the polyhedral program space. %The code of the synthetic program generator is accessible on GitHub at \url{[redacted_link]}

\subsection{Transformation Space Sampling: Relevance-Guided Exploration}
A random sampling of the transformation space is an inefficient method for generating training data. The vast majority of arbitrary transformation sequences are either illegal (i.e., they violate data dependencies) or yield poor performance. A dataset filled with such irrelevant examples would be of little value for training a practical performance model. To create a dataset that is useful for training practical models, the sampling process must be biased towards schedules that are not only legal but are also plausible candidates for performance improvement.

To achieve this, we employed a relevance-guided sampling strategy. This approach uses the execution-guided search mechanism of the LOOPer autoscheduler~\cite{looper_pact25} to explore the transformation space for each synthetic program. This mechanism performs a beam search to explore promising sequences of key polyhedral optimizations, including loop fusion, skewing, interchange, reversal, tiling, parallelization, and unrolling.

By using an established search tool to generate the schedules, we ensure that our dataset is populated with examples that a real-world advanced compiler would actually consider. This guided approach focuses the data on the most interesting and challenging parts of the optimization space, making it more effective for training a model to make quality decisions.

Every transformation sequence considered during this process was verified for correctness using standard polyhedral dependency analysis~\cite{vasilache_violated_2006,feautrier_array_1988} to ensure that only legal (i.e., semantics-preserving) schedules were included in the final dataset.

\subsection{Performance Label Generation}
The final stage of the pipeline generates a ground-truth performance label for each \textit{\texttt{(}program\texttt{,} schedule\texttt{)}} pair. Using the Tiramisu compiler framework~\cite{baghdadi_tiramisu_2019}, we applied the specified transformations to generate an executable file for each pair. The resulting object file was then linked with a measurement wrapper and run on our target hardware, a dual-socket system with Intel Xeon E5-2695 v2 processors. To ensure our measurements were stable and robust against system noise, we executed each program version up to 30 times and recorded the full list of execution times. This entire process was a significant computational undertaking, requiring tens of thousands CPU-hours of machine time to generate the full dataset.

While the performance labels in this dataset are specific to this CPU architecture, the work on the LOOPer system \cite{looper_pact25} has shown that the underlying modeling techniques can be adapted and ported to other hardware targets. The dataset is also well-suited for pre-training. A model can be trained on our large dataset to learn the  general relationships between program structures, transformations, and performance. This pre-trained model can serve as a starting point that can be adapted to a new target architecture through transfer learning or fine-tuning, using only a much smaller, more targeted set of measurements from the new hardware. This approach helps mitigate the "cold start" problem that complicates the development of learned optimizers for new architectures.

%==============================================================================
\section{Dataset Diversity and Analysis}
The utility of a dataset for machine learning applications is defined by its scale, the diversity of its examples, and the breadth of its data distribution. This section provides a detailed characterization of \datasetname{} with respect to these properties.

The analysis is presented in two parts. First, Section~\ref{sec:stats} provides a statistical overview of the dataset, summarizing key properties of the programs, their computational workloads, and the distribution of measured performance improvements. We then perform a rigorous quantitative analysis in Second, Section~\ref{sec:diversity} presents a quantitative analysis that compares the programs in \datasetname{} against an existing polyhedral benchmark suite. This comparison is used to evaluate the structural novelty of the dataset and to determine the extent to which it covers a broader structural space than existing benchmarks.

\subsection{High-Level Statistical Landscape}
\label{sec:stats}

\begin{table}[h!]
\footnotesize
\centering
\caption{High-level statistics of the \datasetname{} dataset.}
\label{tab:stats}
\begin{tabular}{ll}
\toprule
\textbf{Metric} & \textbf{Value} \\
\midrule
Total Unique Programs                 & $\sim220$ Thousands \\
Total Schedules Explored               & $\sim28$ Millions \\
Median Schedules per Program          & 70 \\
Data Generation Effort                & $\sim71$ Thousand CPU Hours \\
Speedup Range  & $0.0004\times$ -- 1230$\times$ \\
\bottomrule
\end{tabular}
\end{table}
To characterize the scale and breadth of \datasetname{}, we first present its high-level statistics. The dataset was produced through a large-scale computational effort designed to capture a wide variety of programs and performance outcomes. Table~\ref{tab:stats} summarizes the primary metrics of the full 28-million-point dataset, illustrating its significant size and the extensive range of performance effects captured. The speedup range, spanning from severe slowdowns to substantial accelerations, highlights the dataset's value in training models to distinguish between beneficial and detrimental optimizations.

To provide a deeper view into the dataset's composition, we present a detailed analysis of its statistical properties. These distributions are fundamental to understanding the diversity encapsulated within the dataset and its suitability for training general-purpose, robust machine learning models for compilation.

\begin{figure}[t]
    \centering
    \includegraphics[width=\textwidth]{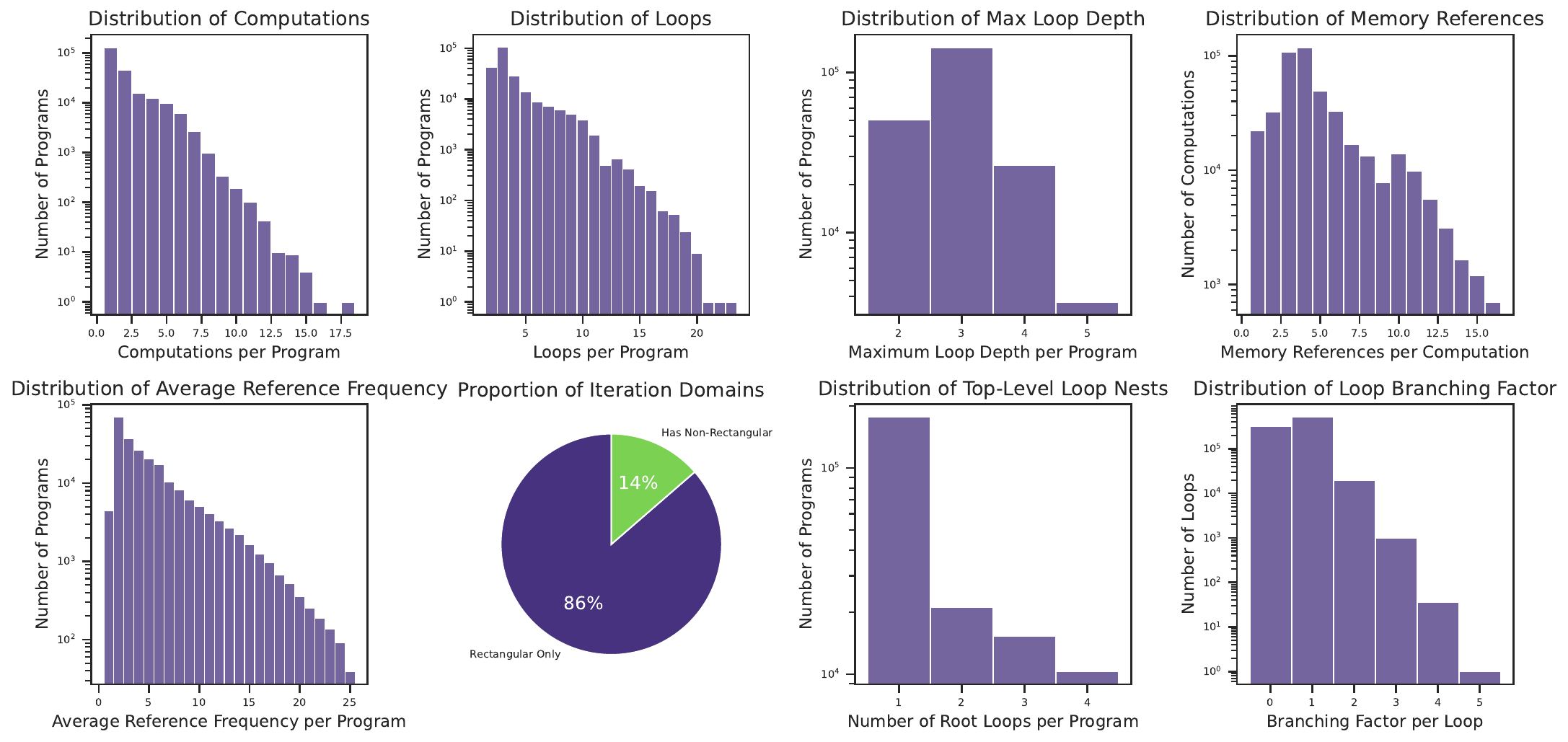}   
    \caption{Distributions of key structural characteristics of the 220k synthetic programs in \datasetname{}. The plots show the diversity in program size, depth, memory access patterns, and iteration domain shapes.}
    \label{fig:structural_diversity} 
\end{figure}

Figure~\ref{fig:structural_diversity} illustrates the structural variety of the synthetic programs. The distributions for the number of computations, loops, and memory references are skewed towards smaller values, reflecting common real-world program sizes, while the long tails ensure that more complex programs are also well-represented. We observe a similar trend in loop depth, memory reference patterns, and branching factors, ensuring that the model is trained on a spectrum from simple, shallow nests to deeper, more intricate configurations. Importantly, 14\% of the programs contain at least one non-rectangular iteration domain, providing vital data for training models that can handle this important class of programs. The distribution of top-level loops indicates that the dataset contains a healthy mix of both single monolithic loop nests and sequences of multiple nests, which is critical for learning about inter-loop optimizations like fusion.

\begin{figure}[t]
    \centering
    
    \includegraphics[width=\textwidth]{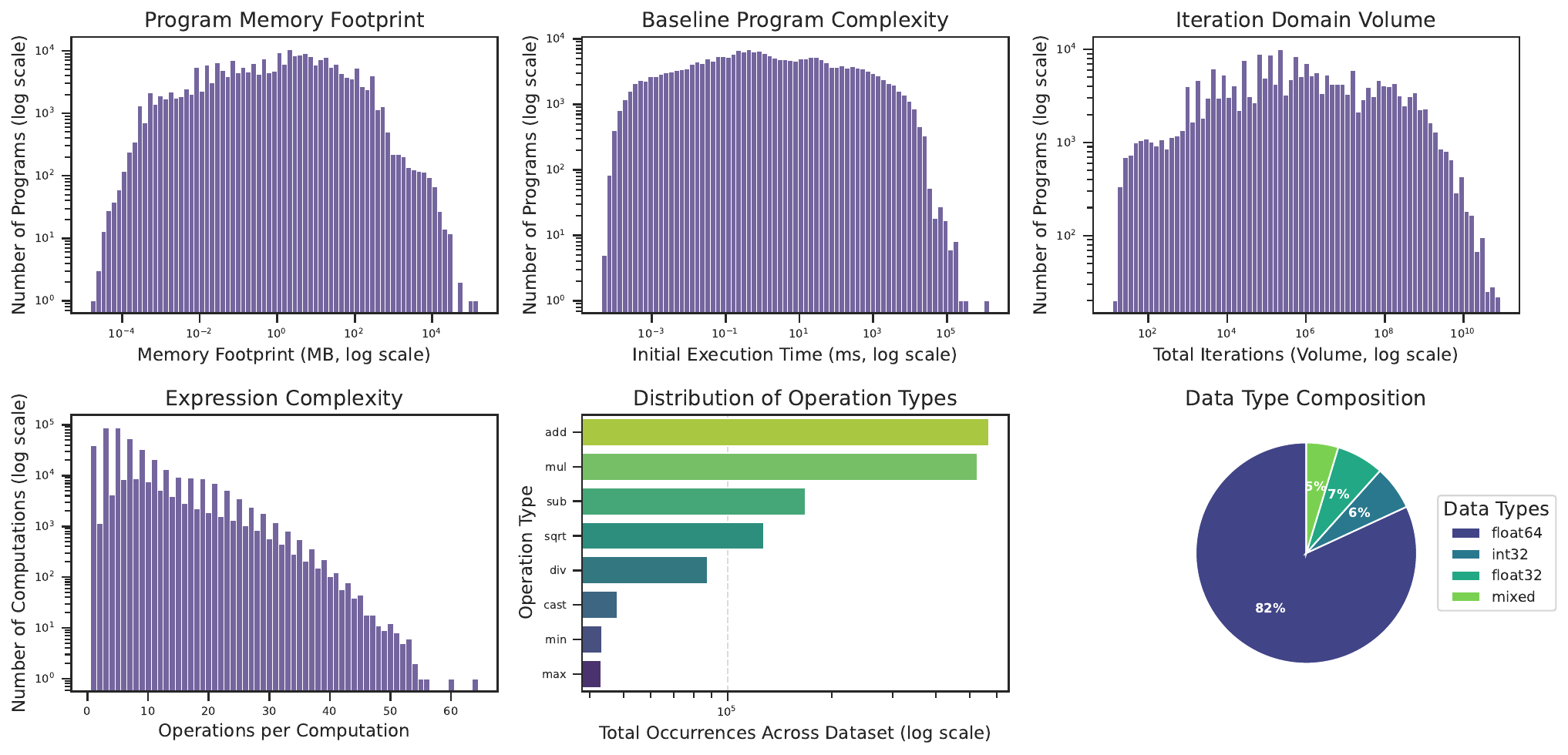}
    \caption{Analysis of program workload and resource consumption. These plots illustrate the dynamic range covered by the dataset in terms of memory footprint, baseline execution time, computational complexity, and low-level operation and type mix.}
    \label{fig:workload_diversity}
\end{figure}

Figure~\ref{fig:workload_diversity} characterizes the computational and memory demands of the programs. The distributions for memory footprint, baseline execution time, and total iteration domain volume all follow a log-normal pattern, spanning many orders of magnitude. This ensures the dataset captures a vast range of program scales, from trivial kernels to computationally expensive ones. Similarly, the complexity of expressions is varied, with most computations being simple but with a long tail of more intensive arithmetic. The distribution of operation types is dominated by common integer and floating-point arithmetic (\texttt{add}, \texttt{sub}, \texttt{mul}), while the data types are primarily \texttt{float64} (82\%), reflecting typical usage in high-performance computing.

\begin{figure}[t]
    \centering
    \includegraphics[width=0.9\textwidth]{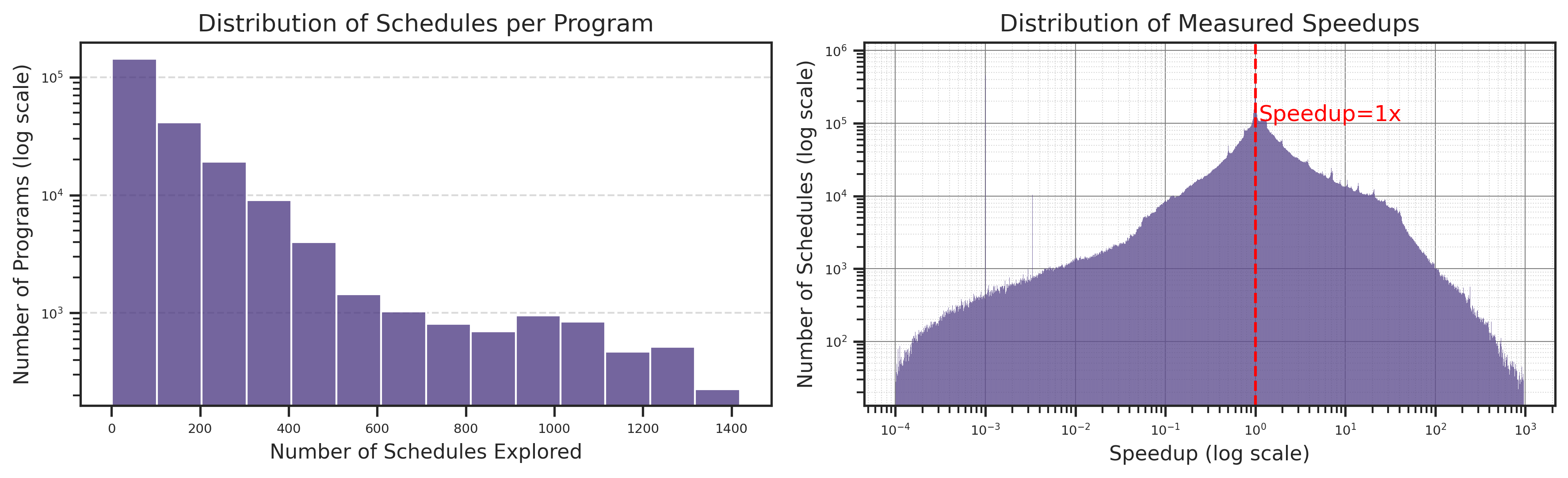}
    \caption{Distribution of the transformation space exploration and the resulting performance impact. The plots show the number of legal schedules explored per program and the wide distribution of measured speedups, which provides the necessary signal for training a performance model.}
    \label{fig:speedup_distribution}
\end{figure}

Finally, Figure~\ref{fig:speedup_distribution} depicts the outcomes of the transformation space exploration. The number of schedules explored per program is highly skewed, indicating that our relevance-guided sampling is efficient for most programs but can explore deeply (over 1,400 schedules) when the search space is rich. The distribution of measured speedups is sharply peaked around 1.0$\times$, which is expected, as many transformations have a neutral or minor effect. The right tail demonstrates that the dataset contains numerous examples of highly effective transformation sequences, while the left tail captures detrimental schedules. This wide and realistic performance distribution is essential for training a robust cost model that can both identify profitable optimizations and avoid costly ones.

\subsection{Quantitative Diversity Analysis}
\label{sec:diversity}
A key question for any synthetic dataset is whether it truly explores new ground or simply re-creates existing examples. To answer this, we performed a quantitative analysis to assess the diversity of our programs and to formally check for any overlap with the PolyBench suite~\cite{pouchet_polybench_2010}.

The results are presented in three parts. We first show that no PolyBench benchmark was accidentally replicated. We then demonstrate that, on the whole, our dataset covers a much broader and more varied structural space. Finally, we provide a complementary view using high-level program features to visualize this diversity.

\subsubsection{Methodology: Measuring Program Similarity}
To quantitatively analyze our dataset's diversity and address the possibility of benchmark replication, we needed a formal way to measure the similarity between any two programs. For this, we chose the normalized Tree Edit Distance (nTED)\cite{doi:10.1137/0218082} as our primary metric .

Our methodology is as follows. First, we represent each program's structure as a tree. The inner nodes of this tree represent loops, and the leaves represent the computational statements within them. To capture the work being done, we then expand each computation leaf into its own expression sub-tree, where the nodes are labeled by their specific arithmetic operation (e.g., \texttt{add}, \texttt{mul}, etc.).

The nTED then calculates the "cost" of transforming one program's tree into another's. We use a uniform cost of 1 for each fundamental edit operation: inserting, deleting, or relabeling a node. To make the comparison fair for programs of different sizes, this raw score is normalized by the total number of nodes in both trees. The final nTED score is a value between 0, for structurally identical programs, and 1, for completely different ones.

It is important to note that this metric focuses purely on the core loop and computational structure. We intentionally omitted details like specific loop bounds, memory access patterns, and data dependencies. This makes our analysis conservative; any measured distance is a lower bound. If we were to include these other distinguishing features, the calculated distances between programs would only increase, further reinforcing our findings about the dataset's diversity.

\subsubsection{Diversity Analysis and Results} 
\begin{figure}[]
    \centering
    \includegraphics[width=\textwidth]{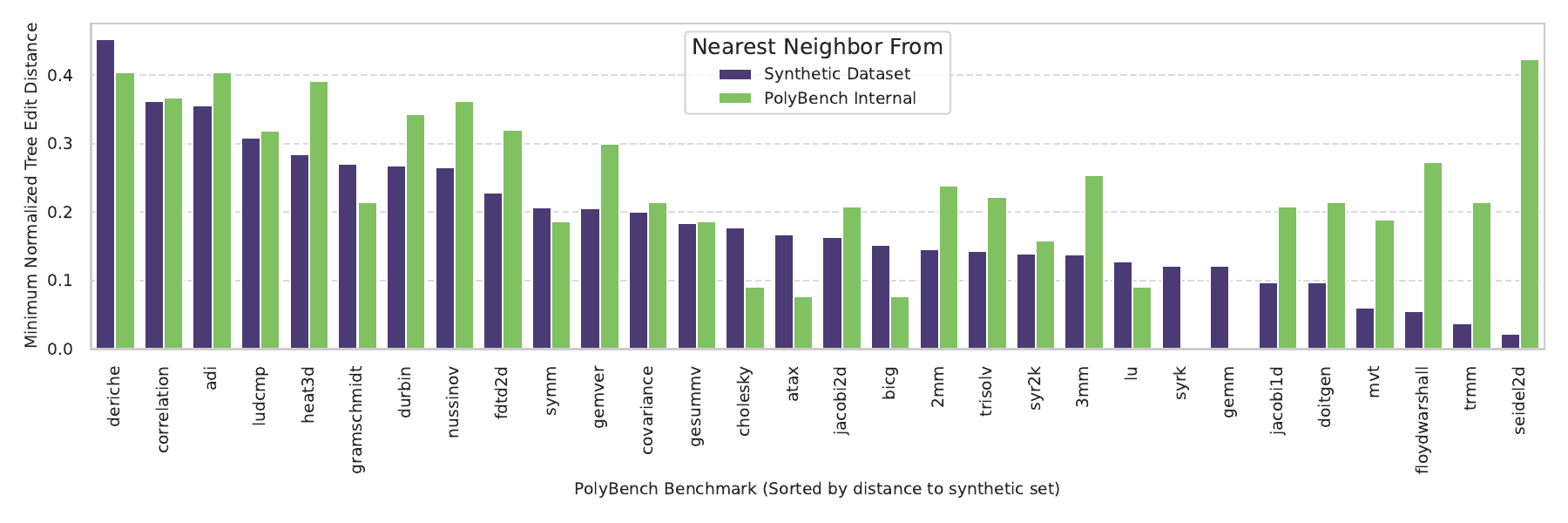} 
    \caption{
        This figure compares the minimum normalized Tree Edit Distance from each PolyBench benchmark to our 220,000-program synthetic dataset (purple) and to the other benchmarks within the PolyBench suite (green). The minimum distance to a synthetic program is never zero, confirming the absence of direct replication. %The intra-benchmark comparison correctly identifies the zero-distance match between \texttt{gemm} and \texttt{syrk}, validating our metric.
    }
    \label{fig:min_ted_barchart}
\end{figure}
\paragraph{Analysis of Minimum Distance:}
Our first and most critical goal was to confirm that we did not accidentally replicate any of the PolyBench benchmarks. Figure~\ref{fig:min_ted_barchart} provides the direct evidence. For each benchmark in PolyBench, it shows the structural distance to the single closest program it has in our entire 220K-program synthetic dataset.

The results are clear: the minimum distance is never zero. The most similar synthetic program we found is to \texttt{seidel-2d}, and even it has a structural distance (nTED) of 0.022. This confirms that no program in PolyBench has an exact structural counterpart in \datasetname{}, even when using our conservative, structure-only metric.

To build confidence in this finding, the figure also includes a baseline: the minimum distance from each benchmark to all other benchmarks within the PolyBench suite itself. Here, our metric performs exactly as expected. It correctly identifies that \texttt{gemm} and \texttt{syrk} (two benchmarks known to share the same core loop and expression structure and differ in iteration domain shape and memory accesses) have a distance of 0.0. Since our metric can correctly find identical structures when they exist, its failure to find any between our dataset and PolyBench provides strong evidence that our generator successfully avoided direct replication.

\begin{figure}[]
    \centering
    \includegraphics[width=0.9\textwidth]{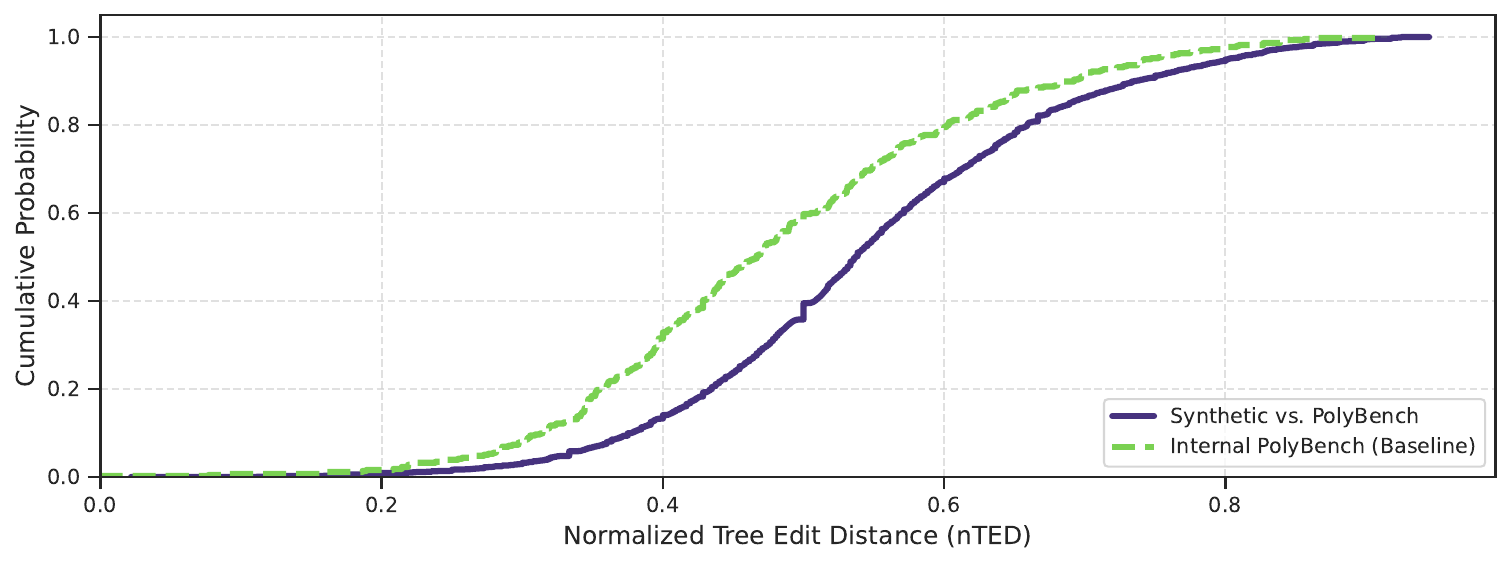} 
    \caption{
        This ECDF plot contrasts the distribution of nTEDs between our synthetic programs and PolyBench (purple) with the distribution of distances within the PolyBench suite itself (green). %The \textit{Synth-vs-Bench} curve's consistent position below the \textit{Bench-vs-Bench} curve demonstrates that our dataset explores a broader and more diverse structural space.
    }
    \label{fig:ecdf_ted}
\end{figure}
\paragraph{Analysis of Distance Distributions:}
Having established that we did not replicate any benchmarks, we next sought to assess if our dataset explores a broader structural space than PolyBench itself. Figure~\ref{fig:ecdf_ted} addresses this by comparing the overall distribution of program distances.

The plot shows the cumulative distribution for two scenarios: the distances between our synthetic programs and the benchmarks (in purple), and the distances between the benchmarks themselves (in green). The visual gap between the two curves is significant. The purple curve is positioned consistently below the green one, which means that for any given level of similarity, a smaller fraction of our synthetic programs are that similar to a benchmark. In simpler terms, it shows that a synthetic program chosen at random is statistically more different from a benchmark than two benchmarks are from each other.

This visual trend is confirmed by the numbers. The median structural distance (the 50th percentile) between our synthetic programs and the benchmarks is 0.537. This is notably higher than the median distance of 0.467 within the PolyBench suite itself.

This result demonstrates that \datasetname{} is not simply a collection of programs that cluster around the structures found in PolyBench. Instead, it actively explores a wider and more varied structural space, making it a more comprehensive resource for training code optimization models.

\begin{figure}[]
    \centering
    \includegraphics[width=\textwidth]{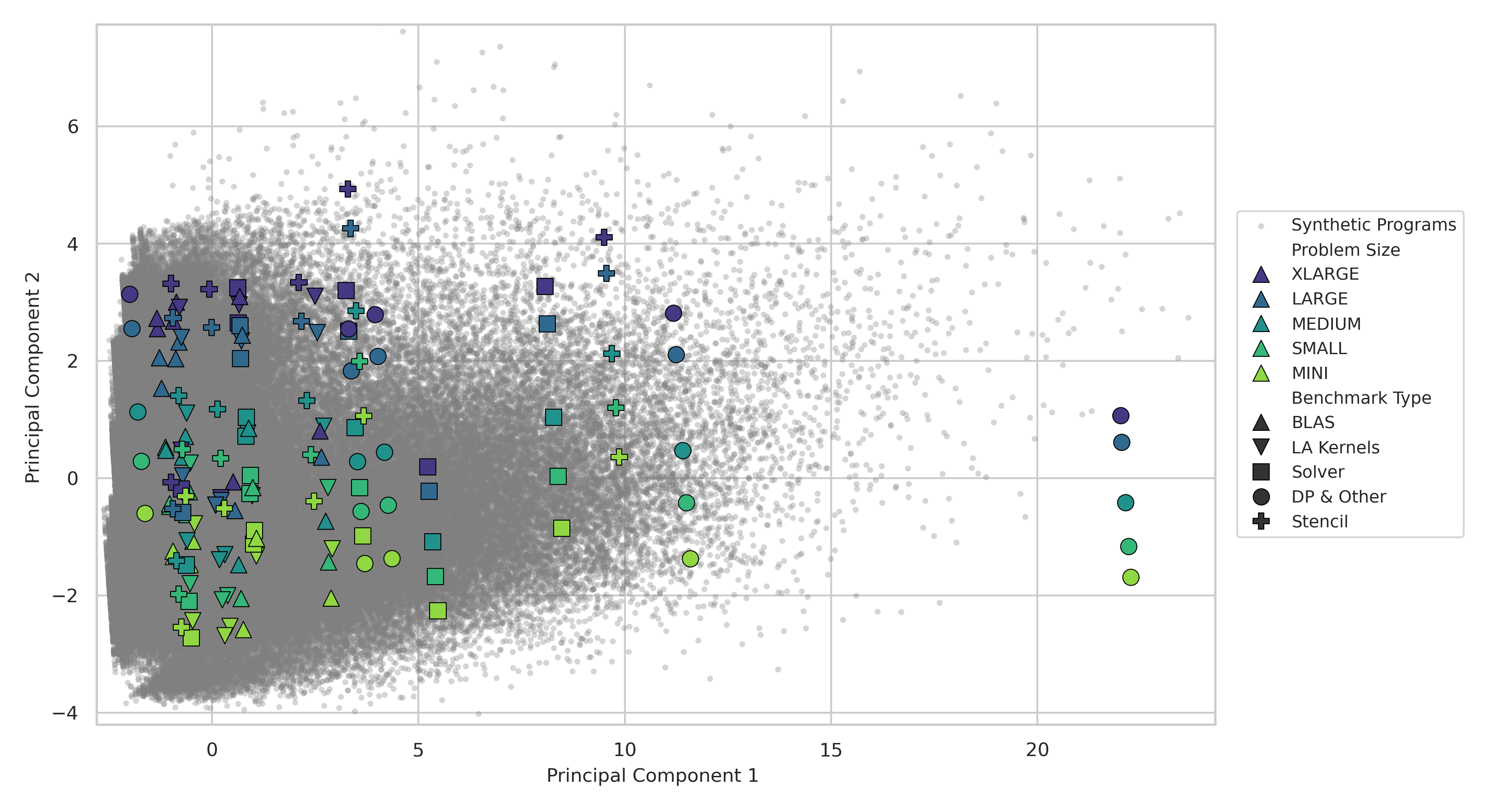} 
    \caption{
    Feature space visualization of \datasetname{} and PolyBench.
        This plot shows a 2D PCA projection of the 20-dimensional feature vectors for our 220,000 synthetic programs (grey cloud) and the 30 PolyBench benchmarks (colored markers). %PolyBench is clearly embedded within the densest regions of our dataset, illustrating that it represents a small, well-covered subset of the much larger feature space captured by \datasetname{}.
    }
    \label{fig:pca}
\end{figure}
\paragraph{A Feature-Based View of Diversity}
To get a different perspective on our dataset's diversity, we moved from the detailed structural comparison of Tree Edit Distance to a higher-level view based on program features. We first represented each program with a vector of 20 key features capturing its size, workload, and memory characteristics (e.g., loop count, memory usage, operation mix).

Since it's impossible to visualize a 20-dimensional space, we used Principal Component Analysis (PCA) to project this data down into the two dimensions that capture the most variance. The result is a "map" of our dataset, shown in Figure~\ref{fig:pca}.

The plot reveals that our 220K synthetic programs form a vast, dense cloud, indicating that our generator produced programs with a wide and continuous range of features. When we overlay the PolyBench programs onto this map, we see that they fall squarely within the densest parts of our synthetic data cloud. This provides a powerful visual confirmation: the types of programs represented by PolyBench are a well-covered, but very small, subset of the much larger feature space captured by \datasetname{}.

We also note that the \texttt{deriche} benchmark appears at the edge of the main cloud, highlighting it as a structurally unique program. This points to a potential area for even further dataset expansion, but overall, the analysis confirms our dataset's comprehensive coverage of common program types.

%==============================================================================
\section{Potential Applications and Research Directions}
While \datasetname{} was created to train the LOOPer system, its value extends to the broader compiler research community as a foundational resource for several key research directions.

The dataset's primary application is to serve as a large-scale benchmark for training and evaluating novel predictive cost models. While the original LOOPer work demonstrated an LSTM-based architecture, \datasetname{} provides the data necessary to explore alternatives, such as Graph Neural Networks (GNNs) that may better capture program dependence graphs, or Transformers designed to identify long-range patterns in optimization sequences. Furthermore, it offers a stable testbed for designing and validating new methods of program representation and featurization, a critical aspect of effective code optimization models.

Beyond model training, \datasetname{} enables the direct empirical analysis of the relationship between transformations and performance. The data can be mined to discover novel, data-driven compiler heuristics, potentially improving the logic of traditional, non-learning-based optimizers. Conversely, it provides a valuable resource for model interpretability and robustness analysis. By identifying classes of programs or optimization schedules where models consistently fail, researchers can diagnose systematic blind spots, guiding the development of more generalizable models.

A central challenge in learned compilation is hardware portability, and \datasetname{} provides a foundation for addressing this problem. The dataset is well-suited for the pre-training of general performance models, which can then be adapted to a new hardware target via transfer learning and fine-tuning on a much smaller, target-specific set of samples. This methodology can significantly reduce the data generation cost associated with porting a learned optimizer, helping to mitigate the "cold start" problem. Furthermore, the dataset's synthetic programs provide a consistent basis for the community to generate new performance labels on diverse hardware, such as GPUs or other CPU microarchitectures, to create a powerful cross-architecture resource for studying zero-shot or few-shot generalization.

%==============================================================================
\section{Format and Accessibility}
\label{sec:format}

We have made the \datasetname{} dataset publicly accessible to ensure that it can be easily used by the research community. This section provides the necessary details on how to access, understand, and use the data.

\subsection{Availability and Access}

The dataset is hosted on the Hugging Face Hub, a platform chosen for its ease of access and its integration with modern data science tools. Two versions of the dataset are available:

\begin{itemize}
    \item The full 28-million-point dataset, containing all programs and schedules generated during our research.
    \item A smaller 10-million-point version, which is pre-split into a \texttt{training} set and a \texttt{validation} set. This split is identical to the one used to train and evaluate the LOOPer cost model~\cite{looper_pact25}, enabling direct replication of its main results.
\end{itemize}

The dataset can be accessed at: \url{https://huggingface.co/datasets/Mascinissa/LOOPerSet}

\subsection{Data Format and Structure}

The dataset is provided in the JSON Lines (\texttt{.jsonl}) format. This is a simple and standard text format where each line is a  self-contained JSON object. Each JSON object represents one unique synthetic program and contains all the optimization schedules that were explored for it.

At a high level, each program's JSON object is structured with the following key fields:
\begin{itemize}
    \item \texttt{program\_name}: A unique identifier for the synthetic program.
    \item \texttt{program\_annotation}: A detailed, structured representation of the original, untransformed program. This includes its loop structure, memory access patterns, and computational expressions, and serves as the primary source for feature engineering.
    \item \texttt{initial\_execution\_time}: The baseline performance of the original program before any optimizations.
    \item \texttt{schedules\_list}: A list of all optimization sequences explored for this program. Each entry in the list details the transformations that were applied and records the final measured execution time, providing the performance labels for model training.
\end{itemize}

An example snippet showing the structure of a single datapoint is provided in Listing~\ref{lst:json_example}. For a complete and detailed specification of every field and sub-field, please refer to the \texttt{README.md} file included with the dataset on Hugging Face.

\begin{figure}[h!]
\begin{lstlisting}[language=json, style=jsonstyle, caption={A simplified example of a single data point in the JSONL format. Each line in the dataset file is a similar JSON object, containing one program and all its measured schedule variants.}, label={lst:json_example}]
{
  "program_name": "function12345",
  "program_annotation": {
    "memory_size": 4.19, // in MB
    "iterators": { ... }, // Contains full loop nest hierarchy
    "computations": { ... }, // Contains expressions and memory accesses
    "buffers": { ... }
  },
  "initial_execution_time": 1393.751, //in ms
  "schedules_list": [
    {
      "execution_times": [451.234, 465.112, 458.543, ...], // List of 30 runs
      "sched_str": "F({C0,C1},1)T2({C0},L2,L3,32,32)...", // Human-readable summary
      "fusions": [["comp00", "comp01", 1]],
      "comp00": {
        "tiling": {"tiling_depth": 2, "tiling_dims": ["L2", "L3"], "tiling_factors": [32, 32]},
        "parallelized_dim": Null,
        ...
      },
      "comp01": { /* ... similar transformation details ... */ }
    },
    { /* ... another schedule object ... */ }
  ]
}
\end{lstlisting}
\end{figure}

\subsection{Tooling and Usage}

Because the data is in the standard JSONL format, it can be loaded and processed easily using any modern programming language and a standard JSON library. For example, loading the dataset in Python is straightforward:

\begin{lstlisting}[language=Python, style=jsonstyle, caption={A simple Python script for loading the dataset.}]
import json

dataset = []
with open('looper_data.jsonl', 'r') as f:
    for line in f:
        dataset.append(json.loads(line))
\end{lstlisting}

For more advanced use, such as constructing the specific input features used by the LOOPer cost model, the original featurization code is publicly available. This code demonstrates how to process the detailed \texttt{program\_annotation} and schedule information to create the input vectors for a machine learning model. It can be found in the official LOOPer Cost Model repository. %at: \url{[Link_will_be_provided_upon_publication]}

\subsection{License}

The \datasetname{} dataset is released under the Creative Commons Attribution 4.0 International (CC-BY 4.0) license. This permissive license allows for the free use, sharing, and adaptation of the dataset for any purpose, including commercial research, provided that appropriate credit is given to the original authors by citing this paper.

%==============================================================================
\section{Conclusion}
This paper introduced \datasetname{}, a large-scale public dataset for data-driven polyhedral compilation. With over 28 million labeled data points from 220,000 unique synthetic programs, the dataset addresses the challenge of data scarcity that often impedes research in this area. By providing a publicly available corpus of programs, transformation schedules, and their corresponding empirical performance, \datasetname{} is intended to lower the barrier to entry and facilitate more direct and reproducible comparisons of learned optimization techniques.

The dataset was constructed using a methodology designed to ensure both structural diversity and practical relevance. This was achieved through a multi-stage synthetic program generator and a relevance-guided strategy for sampling the transformation space. A quantitative analysis using Tree Edit Distance confirmed that the resulting dataset is internally diverse and covers a significantly broader structural space than existing benchmark suites such as PolyBench.

By reducing the high initial cost of data acquisition, \datasetname{} enables researchers to focus on core challenges in machine learning for compilers. It provides a large-scale testbed for developing new model architectures, designing more effective program featurization techniques, and conducting controlled studies on hardware portability and transfer learning. We hope that this resource will support the community in advancing the state-of-the-art in automated code optimization.

\section*{Acknowledgments}
This research was partly supported by the Center for Artificial Intelligence and Robotics (CAIR) at New York University Abu Dhabi, funded by Tamkeen under the NYUAD Research Institute Award CG010. The authors also wish to thank the Commit research group, led by Professor Saman Amarasinghe in the Computer Science and Artificial Intelligence Laboratory (CSAIL) at MIT, for providing access to their computing cluster for the data generation process.

%==============================================================================
%   REFERENCES
%==============================================================================
\bibliographystyle{plain}
\bibliography{references}

\end{document}